# Nuclear Waste Transmutation in Subcritical Reactors Driven by Target-Distributed Accelerators

Anatoly Blanovsky
*Teacher Technology Center, 7850 Melrose Ave., Los Angeles, CA 90046, USA*
*E-mail* ablanovs@lausd.k12.ca.us

**Abstract-***A radioactive waste transmutation system based extensively on existing nuclear power technology is presented. By replacing the control rods with neutron generators, we could maintain good power distribution and perform long-lived waste burning in high flux subcritical reactors (HFSR). To increase neutron source intensity the HFSR is divided into two zones: a booster and a blanket. A neutron gate (absorber and moderator) imposed between two zones permits fast neutrons from the booster to flow to the blanket. Neutrons moving in the reverse direction are moderated and absorbed. The design is based on a small pressurized water reactor (PWR), fission electric cell (FEC), target-distributed accelerator (TDA) and power monitoring system with in-core gamma-ray detectors, now under development in several countries. The FEC is essentially a high-voltage power source that directly converts the kinetic energy of the fission fragments into electrical potential of about 2MV. The TDA, in which an FEC electric field compensates for lost beam energy in the target, offers a new approach to obtain large neutron fluxes.*

## INTRODUCTION

In the conventional reactor, nuclear energy (electrical in nature) is converted to thermal, then to mechanical and then to electrical energy. To achieve the high degree of the burn-up, only fresh low enriched uranium fuel is used. As the surplus reactivity is large, a large amount of the burnable poison material has to be put in the conventional reactor at the expense of the neutron economy.

A typical fuel assembly remains in the PWR for 3 years to a total burn-up of thirty thousand MWD/ton. The burn-up limitation is mainly because of criticality, but not due to radiation damage to the fuel elements. The burn-up for a total of sixty thousand MWD/ton is possible with conventional zircaloy clad fuel elements.

The majority of the fission wastes have a half-life of less than one year. However, some fission waste products, as well as actinides, need the long-term storage. The loss of about 99% of the fuel energy content and concern about safety of the burial solution justify alternative studies. One such alternative is the HFSR with circulating fuel. The HFSR module could be placed in the PWR-type pressure vessel, and a traditional reactor system could be used to pump coolant to steam generators. The HFSR internal separators include heat exchangers for heat transfer and some fission product deposition[1].

## I. HFSR CONCEPT

With self-amplifying due to feedback, blanket and booster multiplication factors of k=0.95 and 0.98, respectively, an external neutron source rate of at least $10^{15}$ n/s is needed to drive the HFSR that produces 300MWt. Most of this power could be generated in the blanket with solid and liquid fuel. Use of a liquid actinide fuel permits transport of the delayed-neutron emitters from the blanket to the booster where they can provide additional neutrons or all the necessary excitation without an external neutron source[2].

Managing the waste is significantly influenced by fission of the actinides (primarily Np-237 and Am-241). The most effective way using the HFSR is to burn the actinides dissolved in water or in molten salt. Since Np-237 and Am-241 fission cross-sections are relatively low they would serve as an absorber in low neutron flux of the dense absorber zone. In high thermal neutron flux of the blanket, they would serve as a fuel. In this design, the plutonium-neptunium fuel circulates in the blanket and the booster; the americium-curium fuel circulates in the outer reflector and absorber zone. A preliminary analysis indicates that an average thermal neutron flux of about $10^{15}$ n/cm$^2$s is achievable in the blanket with an average power density of 300 W/cm$^3$ and tight heavy water lattice[3].

I.A. Actinide Transmutation

The HFSR is mainly a neutron amplifier controlled by external neutron source intensity (source-dominated mode) or by variable feedback (self-amplifying mode). The self-amplifying HFSR is a quasi-critical system in which the blanket flux depends on its power density. The effect of the fuel circulation on the HFSR reactivity is in the increasing role of delayed neutrons in the blanket. The HFSR module includes the distributed target, the booster and the blanket that consists of vented fuel assemblies arranged in annular rings. In order to prevent the loss of neutrons, the reflectors surround the booster and the blanket (see Fig.1).

The design considered here has a fuel assembly configuration utilizing the flux trap principle. The modified PWR fuel assemblies include a central zone with several channels containing liquid actinide fuel. It functions as a flux trap area with high thermal neutron flux. A zone of depleted uranium and/or thorium-based fuel surrounds the central zone. Removing gas and volatile precursors of fission products with high thermal cross sections in the internal separators could reduce a neutron poison.

The actinide fractional loading is optimized to control reactivity, flatten power distribution and produce a partial isotopic separation of nuclides. To achieve a significant actinide burning efficiency, the volume ratio of the heavy water moderator to fuel in the central region of the fuel assembly is high as possible. In high thermal neutron flux, the number of neutrons required for destruction of Np-237 and Am-241 by fission is substantially less than if decay allowed Np-238 and Am-242 to proceed farther up in the neutron capture chain[4].

In the central zone, the Pu-239 content will be much reduced because physical separation of U-238 and actinides. Plutonium with the high Pu-241 content that has an extremely high value of eta at epithermal energy and the delayed-neutron emitters are then continually transported into the booster. To convert U-238 or thorium into fissionable nuclei, the moderator/fuel volume ratio in the depleted fuel region is considerably lower than that in the liquid actinide fuel zone. To satisfy thermal hydraulic constraints the moderator/fuel volume ratio in the depleted fuel region is in the range of 1.0-2.0.

By increasing depleted fuel loading from zero up to 2000 kg, the blanket can be designed to have desired effective enrichment and radius. To have power density in the desired range, the blanket volume is about $10^6 cm^3$. Photoneutrons produced in the blanket play an important part in the actinide transmutation rate (about 100 kg/year).

As the plutonium inventory in the depleted zone increases, it compensates for fissile fuel consumption. At fissile fuel loading of about 5 kg, the blanket would have k=0.95 up to maximum burn-up. After a few fuel cycles with a fertile fuel replacement, the next feed for actinide transmutation is prepared by removing the cladding metal and separation of the uranium. A Purex process can be used to separate neptunium and a Truex process to separate americium and curium from bulk waste.

Long-lived radioactive waste transmutation could be based on a microwave separator, in which a feed material consisting of mixed isotopes is ionized at first stage by electron cyclotron resonance. The ionized material is then fed to an ion cyclotron resonance unit, which is tuned to preferentially excite the minor species. After actinide and long-lived radioactive waste transmutation, direct conversion of radioactive decay energy of short-lived waste to electrical energy can be performed.

I.B. HFSR Control System

Feedback signals from the FEC arrays and calorimetric gamma-ray detectors could be used to control the power of the accelerator sections. This concept of in-core control system was tested at the VVER-440 of the Armenian nuclear power plant. The power distribution and thermal state of the core were computed every 20 seconds on the basis of 2K signals. The time-dependent power-to-signal conversion factor was determined by simple recurrent formula from the previous values. Off-line calculations were used for real-time synthesis of the signals into 3-D power distribution[5].

As the relaxation length of the power release is $L/(1-k)^{1/2}$, where L is the neutron migration length, the HFSR axial and radial distributions is similar to a commercial PWR. Since the control can be basically performed electronically without use of boron chemical control and slow complicated control rods, this simplifies core control and improves neutron economy. Conventional poison rods can be only employed for emergency shutdown of the reactor.

The HFSR control system could also include a digital reactivity meter for subcriticality monitoring and a TV satellite dish with a built-in GPS receiver to analyze the ultrasonic concept the plant seismic resistance increase. To monitoring crustal deformation of the earth, the system that partially based on U.S. patent #6247367 requires radio signals received by antennas on a long baseline. As such motions are preceded an earthquake, a GPS-based earthquake prediction technique could be developed.

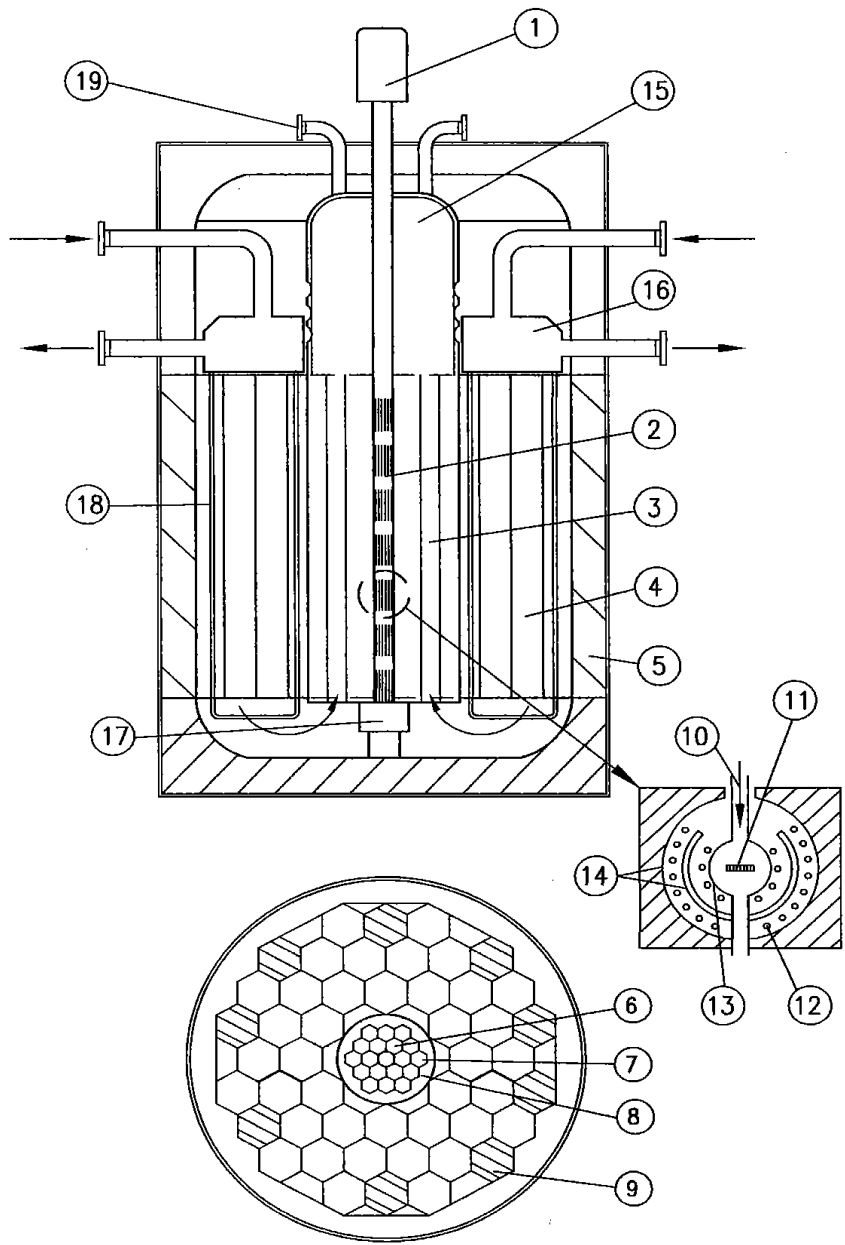

FIGURE 1. SCHEMATIC LAYOUT OF THE HFSR MODULE.

1. Accelerator; 2. Target; 3. Booster; 4. Blanket; 5. Reflector; 6. FEC; 7. Enriched Uranium; 8. Moderator; 9. Spent Fuel; 10. Beam; 11. Target; 12. Grid; 13. Cathode; 14. Anode; 15. Inner Vessel; 16. Separator; 17. Load; 18. Loop; 19. Vacuum pipe.

## II.  PHOTONEUTRON TDA

Now the most used photoneutron target is a thick tungsten layer followed by a beryllium layer. As the photon yield has maximum at about 0.3-0.5R, where R is the range of electrons, the target made of a thin high-Z metal, such as tungsten, lead or tantalum, with heavy water cooling could be used. Monte Carlo calculations of the total neutron yield per incident electron (n/e) were made using MCNPX code. With 10 thin lead targets (0.2cm) in a thick lead tube and 30MV post acceleration, n/e is about 0.02 at electron energy of 18MeV.

The electron TDA has a high voltage accelerating tube composed of insulating rings and metal electrodes. To enhance the efficiency the tube module is divided into an active zone and a trapping zone that contains a reverse biased electrode. This comprises a pair of spaced grids between which an annular metal electrode-target is installed. After impacting the thin target, the electrons are post-accelerated by the distributed transformer-rectifier circuits or the FEC.

The distributed transformer-rectifier circuits could include capacitors that are made by placing tantalum foils on an insulating sheet (tantalum oxide). The foils can form radially arranged capacitors in a Cockcroft-Walton voltage multiplier. However, the FEC with monitoring feedback and small internal resistance is inherently high-voltage power source for in-core applications[6].

Target heating should not be a problem since main energy loss by electron beam is radiative. Photoneutrons are mostly generated in tantalum, heavy water and beryllium surrounded photon production targets. The possible candidates of the target materials are uranium salt in heavy water as an electrolyte is a highly condensed plasma, and thorium oxide as it has high dielectric constant in optical range of frequency. Also, the TDA, in which an electric field compensates for lost beam energy in the gas neutron production target, can be used[7].

The FEC array forms a high-voltage power source connected to the TDA. Each FEC has a tube-type thin fuel cathode with ribs and a quasi-spherical or cylindrical anode, nested in a hexagonal moderator. As this design has an extremely low fuel inventory, liquid phase fuel might be used. After the fuel fission, the more massive fission fragments can reach the anode and deposit their charges. If the FEC cathode is used as a part of the target, it could be separated from the electron guide by a metal membrane. Since in-core applications would benefit from more compact accelerators, a relativistic magnetron driven by the FEC can be used as a microwave power sources for the TDA.

Since the fission fragments are naturally separated into two energy groups, a high efficiency and better electrical field distribution can be obtained through the use of a multistage collector (anode). For a two-stage collector, second collector is made opaque. The first collector made of the thin metal ribbons is essentially transparent to the incoming fragments but it is opaque to the fragments that are turned around. This technique gives the highest efficiency of about 60% because of its excellent space-charge handling ability[8].

The charge deposited by the electron beam in the target could be used to suppress the flow of secondary electrons across the gap between the FEC electrodes. As the stopping power is slightly increased, when the charge state of ions increases, the additional electric field across the cathode could increase the FEC electrical efficiency. The possible candidates of the cathode materials are colored alkali halides to extreme the field strength at the cathode.

For a sphere with uniform density of sources S and macroscopic absorption cross-sections $\Sigma_a$, the average flux is $F \approx S/\pi*(\Sigma_a+DB^2)$ and leakage probability $P=B^2L^2/(1+B^2L^2)$. Here the buckling is $B=\pi/R$ and the diffusion mean free path is $L^2=D/\Sigma_a$. Feedback is important in this design. Use of a liquid fuel that transports the delayed-neutron emitters from the blanket could essentially improve the FEC performance. If $D=0.7$cm, $\Sigma_a=10^{-3}$cm$^{-1}$, R=15cm, L=25cm and a fraction of the delayed-neutrons is 0.001, we have $S \approx 10^{12}$ n/s*cm$^3$ and $F \approx 7*10^{12}$ n/s*cm$^2$. The fission fragment current is about 500μA/g for the FEC fuel thickness of 1.5μm.

## III. CHARGED PARTICLE MOTION IN THE TDA

In the TDA, an external electric field acts counter to the stopping power in the target or between the thin targets. As the beam energy E is constant, neutron yield per particle is approximately Y(E)= d/L(E), where d is the thickness of the target. For n thin targets, in which the energy loss is regained by acceleration of the particles between targets, the total neutron yield per particle is nY. The targets except the last one should have a thickness smaller than the range of the particles in the target.

The electron-screening effects that controlled by the number allowed energy states in medium and vacuum could play essential role in this design. If a slab dielectric material fills space between the charged parallel plates, it must set up an electrical displacement and reduce the field in the dielectric. As the plates are connected to a power supply, the potential remains constant. On the other hand, the screening effect will reduce the ionization loss energy of relativistic electrons.

The additional electric field is reduced to $(k-1)/k^2$ of its vacuum value. Here k is the optical dielectric constant of the target material. A radial electric field in the spherical FEC will set up an electrical displacement partially in the dielectric cathode and partially also in the vacuum field medium. The latter is associated with the Maxwell's displacement currents in electromagnetic theory. As metal has a dielectric constant, an electric field gradient can exist in a metal cathode.

There are no other nuclear engineering systems today in which both relativity and quantum electrodynamics have so many applications as accelerator transmutation of waste. However, there is a difficulty with the mathematical formulation theory of electromagnetic wave propagation that is based on the postulate about constancy of the light velocity.

The Lorentz transformation is a mathematical tool that preserves a wave equation in a moving reference system instead of time invariance in Galilean transformation. As it equally applies to any values of the characteristic wave velocity, anisotropy values of light or sound velocity are used in all applications of the Lorentz transformation.

The theory of dispersive waves gives the simplest model of observed phenomena. It leads to the natural introduction of the group velocity and intensity of de Broglie waves into Maxwell's equations[9]. Recently, this wave model and its possible applications were resented at the 5th Sakharov Conference and 7th Wigner Symposium[10].

## CONCLUSION

Although the low energy accelerators are mainly seen as support equipment, recent advances in neutron generator technology have made it possible to use them for energy generation systems. Different neutron producing reactions (D-T, p-Li, γ-n, etc.) can be used. The initial design could be based on the intense D-T neutron sources and circulating-fuel reactor technology. The final design would require analysis of the relativistic effects in the electron TDA, and experiments with the vacuum fission chamber.

## ACKNOWLEDGES

In 1979, it was discovered by members of the Soviet scientific elite that I had the intention to break the code of silence and perhaps emigrate. So I had to leave my work at the Institute for Nuclear Research (Kiev).

Let me say that I am very grateful to my former colleges (Dr. V.A. Libman, Dr. Y.L. Tsoglin, etc.). They helped me to continue my non-classified study and my supervisor (Dr. V. A. Pshenichny) crossed out his name from our joint article in isomeric nucleus study to protest that my name was crossed out from it.

I am very grateful to my colleges at Fairfax Community Adult School for all the good they have done for me. I am also very grateful to Governor Davis for his support, even though the subject of this study is a federal rather than a state issue.